\documentclass[10pt,conference]{IEEEtran}

\usepackage{cite}

\ifCLASSINFOpdf
   \usepackage[pdftex]{graphicx,color}
  
\else
 \fi
\usepackage[cmex10]{amsmath}
\usepackage{amssymb}
\usepackage{bm} 
\usepackage{amsthm}

 \newtheorem{lemma}{Lemma}

 \theoremstyle{remark}

 \newcommand{\x}{\mathbf}
 \renewcommand{\c}{\mathcal}
 \renewcommand{\b}{\mathbb}
 \renewcommand{\t}{\triangleq}
 \newcommand{\LCM}{\mathop{\mathrm{LCM}}\nolimits}

\usepackage{algorithm}
\usepackage{algpseudocode}
\usepackage[caption=false,font=footnotesize]{subfig}
\usepackage{romannum}

\begin{document}
\title{{\fontsize{19}{24}\selectfont Robust Sparse Fourier Transform Based on The Fourier Projection-Slice Theorem}}

\author{\IEEEauthorblockN{Shaogang Wang,  Vishal M. Patel and Athina Petropulu}

\IEEEauthorblockA{Department of Electrical and Computer Engineering\\
Rutgers, the State University of New Jersey, Piscataway, NJ 08854, USA\\
}}
\maketitle
\pagenumbering{gobble}
\thispagestyle{plain}
\pagestyle{plain}

\begin{abstract}

The state-of-the-art automotive radars employ multidimensional discrete Fourier transforms (DFT) in order to estimate various target parameters. The DFT is implemented using the fast Fourier transform (FFT), at  sample and computational complexity of $O(N)$ and $O(N \log N)$, respectively, where $N$ is the number of samples in the signal space. 
We have recently proposed a sparse Fourier transform based on the Fourier projection-slice theorem (FPS-SFT), which applies to multidimensional signals that are sparse in the frequency domain.
FPS-SFT achieves sample complexity of $O(K)$ and computational complexity of $O(K \log K)$ for a multidimensional, $K$-sparse signal. While FPS-SFT considers the ideal scenario, i.e., exactly sparse data that contains on-grid frequencies, in this paper, by extending FPS-SFT into a robust version (RFPS-SFT), we emphasize on addressing noisy signals that contain off-grid frequencies; such signals arise from radar applications. 
This is achieved by employing a windowing technique and a voting-based frequency decoding procedure; the former reduces the frequency leakage of the off-grid frequencies  below the noise level to preserve the sparsity of the signal, while the latter significantly lowers the frequency localization error stemming from the noise. The performance of  the proposed method is demonstrated both theoretically and numerically. 
\end{abstract}

\begin{IEEEkeywords}
Multidimensional signal processing, sparse Fourier transform, automotive radar, Fourier projection-slice theorem.
\end{IEEEkeywords}

\IEEEpeerreviewmaketitle

\section{Introduction}
With the rapid development of the advanced driver-assistance systems (ADAS) and self-driving cars, the automotive radar plays an increasingly important role in providing multidimensional information on the dynamic environment to the control unit of the car.
Traditional automotive radars measure range and range rate (Doppler) of the targets including cars, pedestrians and obstacles using frequency modulation continuous waveform (FMCW). A digital beamforming (DBF) automotive radar \cite{schneider2005automotive}  can provide angular information both in azimuth and elevation \cite{shirakawa20133d} of the targets, which is more desirable in the ADAS and self-driving applications. 

A typical DBF automotive radar uses uniform linear array (ULA) as the receive array. For such configuration and under the narrow-band signal assumption, each radar target can be represented by a $D$-dimensional ($D$-D) complex sinusoid\cite{engels2017advances}, whose frequency in each dimension relates to target parameters, e.g., range, Doppler and direction of arrival (DOA). 
The simultaneous multidimensional parameter estimation of a DBF automotive radar requires intensive processing. The conventional implementation of such processing relies on a $D$-D discrete Fourier transform (DFT), which can be implemented by the fast Fourier transform (FFT). The sample  complexity of the FFT is  $O(N)$, where $N = \prod_{i=0}^{D-1} N_i$ is the number of samples in the $D$-D data cube with $N_i$ the sample length for the $i_{\rm{th}}$ dimension. For $N$ a power of $2$, the computational complexity of the FFT is $O(N \log N)$.
Since $N$ is typically large, the processing via FFT is still demanding for real-time processing with low-cost hardware.


The recently proposed sparse Fourier transform (SFT) \cite{hassanieh2012nearly,gilbert2014recent,pawar2017ffast} leverages the sparsity of signals in the frequency domain to reduce the sample and computational complexity of DFT.  Different versions of the SFT  have been investigated for several applications including  medical imaging, radar signal processing, etc.  \cite{ hassanieh2015fast,wang2017robust}.
In  radar signal processing, the number of radar targets, $K$, is usually much smaller than $N$, which makes the radar signal sparse in the $D$-D frequency domain. Hence, it is tempting to replace the FFT with SFT in order to reduce the complexity of radar signal processing. 
However, most of the SFT algorithms are designed for 1-dimensional ($1$-D) signals and their extension to multidimensional signals are usually not straightforward.
This is because the SFT algorithms are not separable in each dimension since operations such as detection within an SFT algorithm must be considered jointly for all the dimensions \cite{wang2017robust}.

Multidimensional SFT algorithms are investigated in \cite{ghazi2013sample,hassanieh2015fast,shi2014light}; those algorithms share a similar idea, i.e., reduction of a multidimensional DFT into a number of $1$-D DFTs. 
The SFT of \cite{ghazi2013sample} achieves the sample and computational complexity lower bounds of all known SFT algorithms by reducing a $2$-dimensional ($2$-D) DFT into $1$-D DFTs along rows and columns of a data matrix. However, such algorithm requires a very sparse signal (in the frequency domain) whose frequencies are uniformly distributed; such limitation stems from the restriction of applying DFT only along axes of the data matrix, which corresponds  to projecting a $2$-D DFT of the data matrix onto the two axes of such matrix. 
In \cite{hassanieh2015fast,shi2014light}, the multidimensional DFT is implemented via the application of $1$-D DFTs on samples along a few lines of predefined and deterministic slopes.
Although employing lines with various slopes leads to more degrees of freedom in frequency projection of the DFT domain, the limited choice of line slopes in \cite{hassanieh2015fast,shi2014light} is still an obstacle in  addressing less sparse signals. Moreover, the localization of frequencies in \cite{hassanieh2015fast,shi2014light} is not as efficient as that of \cite{ghazi2013sample}, which  employs the phase information of the  $1$-D DFTs and recovers the significant frequencies in a progressive manner. Thus, the SFT algorithms of \cite{hassanieh2015fast,shi2014light} suffer from higher complexity as compared to the SFT of \cite{ghazi2013sample}. 

We have recently proposed FPS-SFT \cite{wang2017fps}, a multidimensional, Fourier projection-slice based SFT, which enjoys low complexity while avoiding the limitations of the aforementioned algorithms, i.e.,  it can handle less sparse data in the frequency domain, with frequencies non-uniformly distributed. 
FPS-SFT uses the low-complexity frequency localization framework of \cite{ghazi2013sample}, and extends the multiple slopes idea of \cite{hassanieh2015fast,shi2014light} by using lines of randomly runtime-generated slopes. The abundance of randomness of line slopes enables large degrees of freedom in frequency projection in FPS-SFT. Thus, less sparse, non-uniformly distributed frequencies can be effectively resolved (see Section \ref{sec:FPS-SFT} for details).     
Employing random lines is not trivial, since the line parameters, including the line length and slope set should be carefully designed to enable an orthogonal and uniform frequency projection (see Lemmas $1$ and $2$ in \cite{wang2017fps}).   
FPS-SFT can be viewed as a low-complexity, Fourier projection-slice approach for signals that are sparse in the frequency domain. In FPS-SFT, the  DFT of a  $1$-D slice of the $D$-D data is the projection of the $D$-D DFT of the data to such line. While the classical Fourier projection-slice based method reconstructs the frequency domain of the signal using interpolation based on  frequency-domain slices, the FPS-SFT aims to reconstruct the signal directly based on  frequency domain projections; this is achieved by leveraging the sparsity of the signal in the frequency domain.

While the FPS-SFT of \cite{wang2017fps} considered the case with exactly sparse data containing frequencies on the grid, in this paper we consider off-grid frequencies. FPS-SFT suffers from the frequency leakage caused by the off-grid frequencies.  Also, we address the noise that is contained in the signal.  The frequency localization procedure of FPS-SFT is prone to error, since such low-complexity localization procedure is based on the so-called OFDM-trick \cite{hassanieh2012nearly}, which is sensitive to noise.    
Addressing these issues makes the FPS-SFT more applicable to realistic radar applications where the radar signal contains off-grid frequencies and noise.  We term this new extension of FPS-SFT algorithm as RFPS-SFT.

The off-grid frequencies are also addressed in  
\cite{wang2017robust}, where we proposed a robust multidimensional SFT algorithm, i.e., RSFT. In RSFT, the computational savings is achieved by folding the input $D$-D data cube into a much smaller data cube, on which a reduced sized $D$-D FFT is applied. Although the RSFT is more computationally efficient as compared to the FFT-based methods, its sample complexity is the same as the FFT-based algorithms. 
Essentially, the high sample complexity of RSFT is due to its two stages of windowing procedures, which are applied to the entire data cube to suppress the frequency leakage.

Inspired by RSFT, the windowing technique is also applied in RFPS-SFT to address the frequency leakage problem caused by the off-grid frequencies. Instead of applying the multidimensional window on the entire data as in RSFT, the window in RFPS-SFT, while still designed for the full-sized data, is only applied on samples along lines, which does not cause overhead in sample complexity. To address the frequency localization problem in FPS-SFT stemming from noise, RFPS-SFT employs a voting-based frequency localization procedure, which significantly lowers the localization error. The performance of RFPS-SFT is demonstrated both theoretically and numerically, and the feasibility of  RFPS-SFT in automotive radar signal processing is shown via simulations.


\smallskip
\noindent \textbf{Notation:} 
We use lower-case (upper-case) bold letters to denote vectors (matrix). $[\cdot]^T$ denotes the transpose of a vector. The $N$-modulo operation is denoted by $[\cdot]_N$. $[S]$ refers to the integer set of $\{0, ... , S-1  \}$. 
The cardinality of set $\mathbb{S}$ is denoted as $|\mathbb{S}|$. The DFT of signal $x$ is denoted by $\hat{x}$. $\|\mathbf{W}\|_1, \|\mathbf{W}\|_2$ are the $l_1$ and $l_2$ norm of matrix $\mathbf{W}$, respectively.

\section{Signal Model and Problem Formulation} \label{sec:signalModel}

We consider the radar configuration that employs an ULA as the receive array. Assume that the ULA has $N_1$ half-wavelength-spaced elements. The radar transmits FMCW waveform with a repetition interval (RI) of $T_p$.
We also assume that there exist $K$ targets in the radar coverage. After de-chirping, sampling and analog-to-digital conversion for both I and Q channels, the received signal within an RI can be expressed as a superposition of $K$ $2$-D complex sinusoids and noise \cite{engels2017advances}, i.e., 

\begin{equation} \label{eq:sigModel}
\begin{split}
r(\mathbf{n}) = y(\mathbf{n})+n(\mathbf{n}) =  \sum_{(a,\bm{\omega}) \in \mathbb{S}} a e^{j \mathbf{n}^T \bm{\omega}} + n(\mathbf{n}),\\ 
\end{split}
\end{equation}
where $\mathbf{n} \triangleq [n_0,n_1]^T \in \mathcal{X} \triangleq [N_0]\times [N_1]$ is the sampling grid and $N_0$ is the number of samples within an RI. $y(\mathbf{n}) \triangleq \sum_{(a,\bm{\omega}) \in \mathbb{S}} a e^{j \mathbf{n}^T \bm{\omega}}$ is the signal part of the received signal; $(a,\bm{\omega})$ represents a $2$-D sinusoid, whose complex amplitude is $a$, and it holds that $0<a_{min}\le |a| \le a_{max}$; the $2$-D frequency $\bm{\omega} \triangleq [\omega_0, \omega_1]^T \in [0, 2\pi)^2$ represents the normalized radian frequencies corresponding to targets' range and DOA, respectively. The set $\mathbb{S}$, with $|\mathbb{S}|=K$ contains all the $2$-D sinusoids. The noise, $n(\mathbf{n})$, is assumed to be i.i.d., circularly symmetric Gaussian, i.e.,  $\mathcal{CN}(0, \sigma_n)$. 
The SNR of a sinusoid with amplitude $a$ is defined as $SNR \triangleq (|a|/\sigma_n)^2$. 

The target's rang $r$, Doppler $f_d$ and DOA $\theta$ relate to $\bm{\omega}$ as  $\omega_0 = 2\pi (2\rho r/c+f_d)/f_s, \omega_1 =\pi  \sin \theta $, 
where $\rho,c,f_s$ are the chirp rate, the speed of wave propagation and sampling frequency, respectively; the chirp rate is defined as the ratio of the signal bandwidth and the RI. Thus, the target parameters are embedded in frequencies $\omega_0, \omega_1$, which can detected in the $2$-D $N_0 \times N_1$-point DFT of $r(\x{n})$ \cite{engels2017advances}, i.e., 
\begin{equation} \label{eq:DFT}
\begin{split}
\hat{r}(\mathbf{m}) & \triangleq  \frac{1}{N} \sum_{\mathbf{n} \in \mathcal{X}} w(\mathbf{n}) r(\mathbf{n}) e^{-j2\pi \left(\frac{m_0 n_0}{N_0} + \frac{m_1 n_1}{N_1}\right) },\\
&= \hat{y}(\mathbf{m})+\hat{n}(\mathbf{m}), \; \mathbf{m} \triangleq [m_0,m_1]^T \in \mathcal{X},
\end{split}
\end{equation}
where $w(n)$ is a 2-D window, introduced to suppress frequency leakage generated by off-grid frequencies; $N = N_0 N_1$; and $\hat{y}(\x{m}),  \hat{n}(\x{m})$ are the DFTs of the windowed $y(\x{n})$ and $n(\x{n})$, respectively.
Assuming that the peak to side-lobe ratio (PSR) of the window is large enough, such that the side-lobe (leakage) of each frequency in $\b{S}$ can be neglected in the DFT domain, then $\hat{y}(\x{m})$ is contributed by a set of $2$-D sinusoids, whose frequencies are on-grid, i.e.,  $\b{S}' \triangleq \{(a, \bm{\omega}) : \bm{\omega} \triangleq [2\pi m_0/N_0, 2\pi m_1/N_1]^T, [m_0,m_1]^T \in \mathcal{X}\}$ with $K<|\b{S}'|<<N$. Note that since the windowing degrades the frequency resolution, each sinusoid in $\b{S}$ is related to a cluster of  sinusoids in $\b{S}'$, which can be estimated from $\hat{r}(u,v)$; next, the estimation of $\b{S}$ can be computed from the estimation of $\b{S}'$ via, for example, the quadratic interpolation method\cite{smith1987parshl}.

The sample domain signal component associated with the window $w(\x{n}), \x{n}\in \c{X}$ and the set of sinusoids, $\b{S}'$, can be expressed as 
\begin{equation} \label{eq:sigModel2}
{x}(\x{n}) \triangleq \sum_{(a,\bm{\omega}) \in \b{S}'} a e^{j 2\pi \left( \frac{m_0 n_0}{N_0}   + \frac{m_1 n_1}{N_1}  \right) },\; [n_0,n_1]^T \in \mathcal{X}.
\end{equation}


The state-of-the-art DBF automotive radars also measure the target Doppler $f_d$ by processing  a $3$-dimensional ($3$-D) data cube generated by $N_2$ consecutive RIs \cite{engels2017advances}. The normalized radian frequency $\omega_2$ in the Doppler dimension relates to the Doppler as $\omega_2 = 2\pi f_d T_p$. The DBF automotive radars that also measure elevation DOA of targets introduce a $4$-th dimension of processing\cite{shirakawa20133d}; the DOA measurement in elevation is similar to that of the azimuth DOA dimension. In those  cases,  the proposed RFPS-SFT algorithm can be naturally extended to multidimensional cases, where the reductions of complexity of the signal processing algorithms are more significant.

\section{The RFPS-SFT Algorithm} \label{sec:RFPS-SFT}

\subsection{FPS-SFT} \label{sec:FPS-SFT}
The FPS-SFT algorithm proposed in \cite{wang2017fps} applies to multidimensional data of arbitrary size that is exactly sparse in the frequency domain. In the $2$-D case, FPS-SFT implements a $2$-D DFT as a series of $1$-DFTs on samples extracted along lines, with each line being parameterized by the random slope parameters $\bm{\alpha} \triangleq [\alpha_0, \alpha_1]^T \in \c{X}$ and delay parameters $\bm{\tau} \triangleq [\tau_0, \tau_1]^T \in \c{X}$. The signal along such line can be expressed as 
\begin{equation} \label{eq:sample_slice}
\begin{split}
&s(\bm{\alpha},\bm{\tau},l) \triangleq x([\alpha_0 l+\tau_0]_{N_0}, [\alpha_1 l+\tau_1]_{N_1})\\  &= \sum_{(a,\bm{\omega})\in \b{S}'} a e^{j 2\pi  \left( \frac {m_0  [\alpha_0 l+\tau_0]_{N_0}}{N_0} + \frac{m_1  [\alpha_1 l+\tau_1]_{N_1}}{N_1} \right)},   l\in [L].
\end{split}
\end{equation}

On taking an $L$-point DFT on (\ref{eq:sample_slice}) w.r.t. $l$, we get
\begin{equation}  \label{eq:hs}
\begin{split}
&\hat{s}(\bm{\alpha},\bm{\tau},m) \triangleq \frac{1}{L} \sum_{l \in [L]}  s(\bm{\alpha},\bm{\tau}, l) e^{-j 2\pi \frac{l m}{ L}} \\
 &= \frac{1}{L} \sum_{(a,\bm{\omega})\in \b{S}'} a e^{j 2\pi \left(\frac{m_0 \tau_0}{N_0} + \frac{m_1 \tau_1}{N_1} \right)} \sum_{l \in [L]} e^{j2\pi l \left(\frac{m_0 \alpha_0}{N_0} + \frac{m_1 \alpha_1}{N_1} - \frac{m}{L}\right)},\\
& m \in [L].
\end{split}
\end{equation}

The line length, $L$, which is the least common multiple (LCM) of $N_0,N_1$ is designed such that the  orthogonality condition for frequency projection is satisfied (see Lemma $1$ of \cite{wang2017fps} for details), i.e., for $m \in [L], [m_0,m_1]^T \in \c{X}$, 
\begin{equation} \label{eq:orthogonal}
\begin{split}
& \hat{f}(m) \triangleq \frac{1}{L} \sum_{l \in [L]} e^{j2\pi l \left(\frac{m_0 \alpha_0}{N_0} + \frac{m_1 \alpha_1}{N_1} - \frac{m}{L}\right)} \in \{0,1\},\\
\end{split}
\end{equation}
then if
\begin{equation} \label{eq:lines}
\left[\frac{m_0 \alpha_0}{N_0} + \frac{m_1 \alpha_1}{N_1} - \frac{m}{L}\right]_1 = 0, [m_0,m_1]^T \in \mathcal{X},
\end{equation}
the $m_{\rm{th}}$ entry of (\ref{eq:hs}) can be simplified as $\hat{s}(\bm{\alpha},\bm{\tau},m) = \sum_{(a,\bm{\omega}) \in \b{S}'} a e^{j 2\pi \left(\frac{m_0 \tau_0}{N_0} + \frac{m_1 \tau_1}{N_1} \right)}$.
The solutions of (\ref{eq:lines}) with respect to ${m}$ lie on a line with slope $-\alpha_0 N_1/(\alpha_1 N_0)$ in the $N_0\times N_1$-point DFT domain (see the proof of Lemma $2$ in \cite{wang2017fps}), i.e., for $m_0,m_1$ satisfying (\ref{eq:lines}), it holds that
\begin{equation}
m_0 = [m_0' + k \alpha_1 L/N_1]_{N_0}, m_1 = [m_1' - k \alpha_0 L/N_0]_{N_1}, k \in \b{Z},
\end{equation}
where $[m_0',m_1']^T \in \c{X}$ is one of the solutions of (\ref{eq:lines}). 

Hence each entry of the $L$-point DFT of the slice taken along a time-domain line with slope $\alpha_1/\alpha_0$ represents a projection of the $2$-D DFT along the line with slope $-\alpha_0 N_1/(\alpha_1 N_0)$, which is orthogonal to the time-domain line. This is closely related to the Fourier projection-slice theorem.   
In fact, FPS-SFT can be viewed as a low-complexity, Fourier projection-slice based multidimensional DFT.
This is achieved by exploring the sparsity nature of the signal in the frequency domain, which is explained in the following. 

Assume that the signal is sparse in the frequency domain, i.e., $|\b{S}'| = O(L)$. Then,  if $|\hat{s}(\bm{\alpha},\bm{\tau}, m)| \neq 0$, with high probability, the $m_{\rm{th}}$ bin is $1$-sparse, i.e., contains the projection of the DFT value from only one significant frequency, and it holds that $\hat{s}(\bm{\alpha},\bm{\tau}, m) = a e^{j 2\pi \left(\frac{m_0 \tau_0}{N_0} + \frac{m_1 \tau_1}{N_1} \right)}, (a, \bm{\omega}) \in \b{S}'$.
In such case, the $2$-D sinusoid, $(a, \bm{\omega})$, can be `decoded' by three lines of the same slope but different offsets. The offsets for the three lines are designed as $\bm{\tau}, \bm{\tau}_0 \triangleq [[\tau_0+1]_{N_0}, \tau_1]^T, \bm{\tau}_1 \triangleq [\tau_0, [\tau_1+1]_{N_1}]^T$, respectively; such design allows for the frequencies to be decoded independently in each dimension. The sinusoid corresponding to the $1$-sparse bin, $m$, can be decoded as
\begin{equation} \label{eq:decoding}
\begin{split}
&m_0 = \left[ \frac{N_0}{2 \pi} \phi\left( \frac{\hat{s}(\bm{\alpha},\bm{\tau}_0,m) }{\hat{s}(\bm{\alpha},\bm{\tau},m) } \right)  \right]_{N_0}, \\
&m_1 = \left[ \frac{N_1}{2 \pi} \phi\left(\frac{\hat{s}(\bm{\alpha},\bm{\tau}_1,m)}{\hat{s}(\bm{\alpha},\bm{\tau},m)}    \right) \right]_{N_1},\\
&a = \hat{s}(\bm{\alpha},\bm{\tau},m) e^{-j2\pi(m_0 \tau_0/N_0 + m_1 \tau_1/N_1)}.
\end{split}
\end{equation}

To recover all the sinusoids in $\b{S}'$,  each iteration of FPS-SFT adopts a random choice of line slope (see Lemma $2$ of \cite{wang2017fps}) and offset. Furthermore, the contribution of the recovered sinusoids 
in  previous iterations is removed to create a  sparser signal. Specifically, assuming that for current iteration, the line slope and offset parameters are selected as $\bm{\alpha},\bm{\tau}$, respectively, the recovered sinusoids are projected into $L$ frequency bins to construct the DFT along the line, $\hat{s}_r(\bm{\alpha},\bm{\tau}, m) \triangleq \sum_{(a,\bm{\omega}) \in \mathcal{I}_{m}} a e^{j 2\pi \left(\frac{m_0 \tau_0}{N_0} + \frac{m_1 \tau_1}{N_1} \right)}$, $m \in [L]$, where $\mathcal{I}_{m}, m \in [L]$ represent the subsets of the recovered sinusoids that relate to the constructed DFT along line via projection, i.e.,  $\mathcal{I}_{m} \triangleq \{(a,\bm{\omega}) : [\frac{m_0 \alpha_0}{N_0} + \frac{m_1 \alpha_1}{N_1} - \frac{m}{L}]_1 = 0, [m_0,m_1]^T \in \c{X} \}, m \in [L]$. Next, the $L$-point inverse DFT (IDFT) is applied on $\hat{s}_r (\bm{\alpha},\bm{\tau},m), m \in [L]$, from which the line, ${s}_r (\bm{\alpha},\bm{\tau},l), l \in [L]$ due to the previously recovered sinusoids are constructed. Subsequently, those constructed line samples are subtracted from the signal samples of the current iteration.

\subsection{RFPS-SFT} \label{sec:DL_SFT_overview}

FPS-SFT \cite{wang2017fps} was developed for data that is exactly sparse in the frequency domain. Also, the frequencies are assumed to be on-grid of the $N_0\times N_1$-point DFT. In the radar application however, the radar signal contains noise. Also,  the discretized frequencies associated with target parameters are typically off-grid. In the following, we propose RFPS-SFT, which employs the windowing technique to reduce the frequency leakage produced by the off-grid frequencies and a voting-based frequency localization to reduce the frequency decoding error due to noise.

\subsubsection{Windowing}
To address the off-grid frequencies, we apply a window $w(\x{n}), \x{n} \in \c{X}$ on the signal of (\ref{eq:sigModel}). The PSR of the window, $\rho_w$, is designed such that the side-lobes of the strongest frequency are  below the noise level, hence the leakage of the significant frequencies can be neglected and the sparsity of the signal in the frequency domain can be preserved to some extend. The following lemma reflects the relationship between $\rho_w$ and the maximum SNR of the signal.

\begin{lemma} \label{le:window}
\textbf{(Window Design):} 
Consider (\ref{eq:DFT}), which is the $N_0 \times N_1$-point DFT of signal of (\ref{eq:sigModel}). Let  $\mathbf{W} \in \mathbb{R}^{N_0\times N_1}$ be the matrix generated by the window function $w(\x{n}), \x{n} \in \c{X}$. The PSR of the window, $\rho_w$, should be designed such that
\begin{equation} \label{eq:rho_w}
\rho_w > \frac{2 \|\mathbf{W}\|_1}{\sqrt{\pi} \|\mathbf{W}\|_2}\sqrt{SNR_{max}},
\end{equation}
Where $SNR_{max} \triangleq a_{max}^2/\sigma_n^2$.  
\end{lemma}

Note that unlike the RSFT that applies windows on the entire data cube, in RFPS-SFT, while the window is still designed for the entire data cube, the windowing is applied only on the sampled locations. Thus, the windowing does not increase the sample and computational complexity of RFPS-SFT.

\subsubsection{Voting-based frequency decoding}
When the signal is approximately sparse, the frequencies decoded by (\ref{eq:decoding}) are not integers. Since we aim to recover the gridded frequencies, i.e., $\b{S}'$ of (\ref{eq:sigModel2}), the recovered frequencies are rounded to the nearest integers. When the SNR is low, the frequency decoding could result into false frequencies; those false frequencies enter the future iterations and generate more false frequencies. To suppress the false frequencies, motivated by the  classical $m$-out-of-$n$ radar signal detector\cite{skolnik1970radar}, RFPS-SFT adopts an $n_d$-out-of-$n_s$ voting procedure in each iteration. Specifically, within each iteration of RFPS-SFT, $n_s$ sub-iterations are applied; each sub-iteration adopts randomly generated line slope and offset parameters and recovers a subset of frequencies, $\mathbb{S}_i, i\in [n_s]$. 
Within those frequency sets, a given recovered frequency  could be recovered by $n$ out of $n_s$ sub-iterations.  For a true significant frequency, $n$ is typically larger than that of a false frequency, thus only those frequencies with $n\ge n_d$ are retained as the recovered frequencies of the current iteration.  When $(n_s, n_d)$ are properly chosen, the false frequencies can be reduced significantly.

\subsubsection{The lower bound of the probability of correct localization and the number of iterations of RFPS-SFT} \label{sec:localization}

The probability of decoding error relates to the SNR, signal sparsity and choice of $(n_s,n_d)$ in RFPS-SFT. In the following, we provide the lower bound for the probability of correct localization of the significant frequencies for each iteration of RFPS-SFT, from which one can derive the number of iterations of RFPS-SFT in order to recover all the significant frequencies of sufficient SNR.


According to Section \ref{sec:signalModel}, a $2$-D sinusoid $(a, \bm{\omega}) \in \b{S}$ of (\ref{eq:sigModel}) is associated with a cluster of $2$-D sinusoids $\b{S}_0 \subseteq \b{S}'$ of (\ref{eq:sigModel2}), whose frequencies are on the grid of the $N_0 \times N_1$-point DFT.  Let's assume that the sinusoid $(a_d, 2\pi [m_0/N_0, m_1/N_1]^T) \in \b{S}_0$ with $[m_0,m_1]^T \in \c{X}$ has the largest absolute amplitude among the sinusoids in $\b{S}_0$. In addition we assume that the SNR of $(a, \bm{\omega})$ is sufficiently high, then the probability of correctly localizing $[m_0,m_1]^T$ in each iteration of RFPS-SFT is lower bounded by
\begin{equation} \label{eq:err_lower_bound}
P_d \triangleq \sum_{n'_d = n_d}^{n_s} \binom{n_s}{n'_d} (P_1 P_w)^{n'_d} (1-P_1 P_w)^{n_s-n'_d},
\end{equation}
where $P_1 \t (1-|\b{S}''|/N)^{N/L-1}$ with $L = \LCM(N_0,N_1)$ is the probability of a sinusoid in $\b{S}''$ being projected to a $1$-sparse bin,  and $\b{S}''$ with $\b{S}'' \subseteq \b{S}'$  represents the remaining  sinusoids to be recovered in the future iterations of RFPS-SFT; $P_w \t (1-P_u)(1-P_v)$ is the lower bound of the probability of correct localization for a $2$-D sinusoid that is projected into a $1$-sparse bin for one sub-iteration of RFPS-SFT; $P_u, P_v$ are the upper bounds of the probability of localization error for the two frequency components, $m_0,m_1$, respectively, which is defined as 
$
P_u  \t \left( \sigma_p (1-f_{|a_n|}(\delta_u)) \right)^2, 
P_v  \t \left( \sigma_p (1-f_{|a_n|}(\delta_v)) \right)^2,
$
where $\delta_u \t a \pi \|\mathbf{W}\|_1  /(2N N_0)$, $\delta_v \t a \pi \|\mathbf{W}\|_1  /(2N N_1)$, with $\mathbf{W} \in \b{R}^{N_0 \times N_1}$ the window that is applied on the data; $\sigma_p$ with $ \frac{1}{2} \le \sigma_p \le \frac{1}{2\pi}$ is the parameter determined by the phases of the error vectors contained in the $1$-sparse bin; $f_{|a_n|}(x)$ is the  cumulative distribution function (CDF)  of the Rayleigh distribution, which is defined as
$
f_{|a_n|}(x) \t 1-e^{-x^2/ (2 \sigma^2_{a_n'}) }, \; x>0,
$
where $\sigma^2_{a_n'} \t  \sigma_n^2 \|\mathbf{W}\|^2_2 /(2NL)$.

Essentially, (\ref{eq:err_lower_bound}) represents the complementary cumulative binomial probability resulted from the $n_d$-out-of-$n_s$ voting procedure, where the success probability of each experiment, i.e., localizing $(a_d, 2\pi [m_0/N_0, m_1/N_1]^T)$ in each sub-iteration of RFPS-SFT is $P_1 P_w$.
When $|\b{S}'|$ is known, (\ref{eq:err_lower_bound}) can be applied to estimate the largest number of iterations (the upper bound) of RFPS-SFT in order to recover all the significant sinusoids in $\b{S}'$ since the least number of recovered sinusoids in each iteration can be estimated by $|\b{S}''| P_d$.

\subsubsection{Complexity analysis} \label{sec:complexity}
The RFPS-SFT executes $T$ iterations; within each iteration, $n_s$ sub-iterations with randomized line parameters are invoked. The samples used in each sub-iteration is $3L$, since three $L$-length lines, with $L=\LCM(N_0,N_1)$  are extracted to decode the two frequency components in the $2$-D case. Hence, the sample complexity of RFPS-SFT is $O(3T n_s L) = O(L)$.

The core processing of RFPS-SFT is the $L$-point DFT, which can be implemented by the FFT with computational complexity of $O(L \log L)$.  In addition to the FFT, each sub-iteration needs to evaluate $O(|\b{S}'|)$ frequencies. Hence the computational complexity of RFPS-SFT is $O(L \log L+|\b{S}'|)$. Assuming that $|\b{S}'| = O (L)$, the sample and computational complexity can be simplified as $O(|\b{S}'|)$ and $O(|\b{S}'| \log |\b{S}'|)$, respectively.  Furthermore, since $K=O(|\b{S'}|)$, the sample and computational complexity of RFPS-SFT can be further simplified as $O(K)$ and $O(K \log K)$, respectively.

\subsubsection{Multidimensional extension} \label{sec:multi_dimension}
The multidimensional extension of RFPS-SFT is straightforward and similar to that of FPS-SFT (See Section $2.3$ of \cite{wang2017fps} for details).

\section{Numerical Results} \label{sec:numerical}
\begin{figure}[!htp]
	\begin{center}
	\includegraphics[scale=0.28]{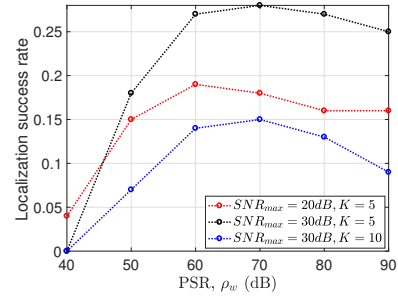} 
        \vskip -10pt
	\caption{Frequency localization success rate of the first iteration of RFPS-SFT versus window PSR. The Dolph-Chebyshev windows with various PSR is applied. $N_0 = N_1 = 256; (n_s,n_d) = (3,2)$. The results are averaged based on $100$ iterations of Monte Carlo simulation.}
	\label{fig:decodingErrPSR}
	\end{center}
\end{figure}

\noindent \textbf{Effect of windowing on frequency localization:}
For the data that contains off-grid frequencies, the PSR of the required window is given in Lemma \ref{le:window}. However, the larger the PSR, the wider the main-lobe of the window, which results into larger frequency clusters in the DFT domain and thus larger $|\b{S}'|$ (see (\ref{eq:sigModel2})), i.e., a less sparse signal. Moreover, the larger the PSR, the smaller the SNR of the windowed signal, which leads to larger frequency localization error. Hence, for a signal with known maximum SNR, $SNR_{max}$, there exists a window with the optimal PSR in terms of frequency localization success rate, i.e., ratio of number of correctly localized frequencies to the number of significant frequencies, which is $|\b{S}'|$ in one iteration of RFPS-SFT.  Fig. \ref{fig:decodingErrPSR} shows the numerical evaluation of such optimal windows for signals for various values of $SNR_{max}$ and sparsity level, i.e., $K = |\b{S}|$. According to (\ref{eq:rho_w}), for signals with $SNR_{max}$ equal to $20dB$ and $30dB$, the PSR of the window should be larger than $56dB$ and $60dB$, respectively. The corresponding optimal PSR for the Dolph-Chebyshev windows appear to be $60 dB$ and $70 dB$, respectively. 
Fig. \ref{fig:decodingErrPSR} shows the success rate of the first iteration of RFPS-SFT, which is the lowest success rate of all the iterations.

\begin{figure}[!htp]
    \centering
    \subfloat[]{{\includegraphics[scale=0.21]{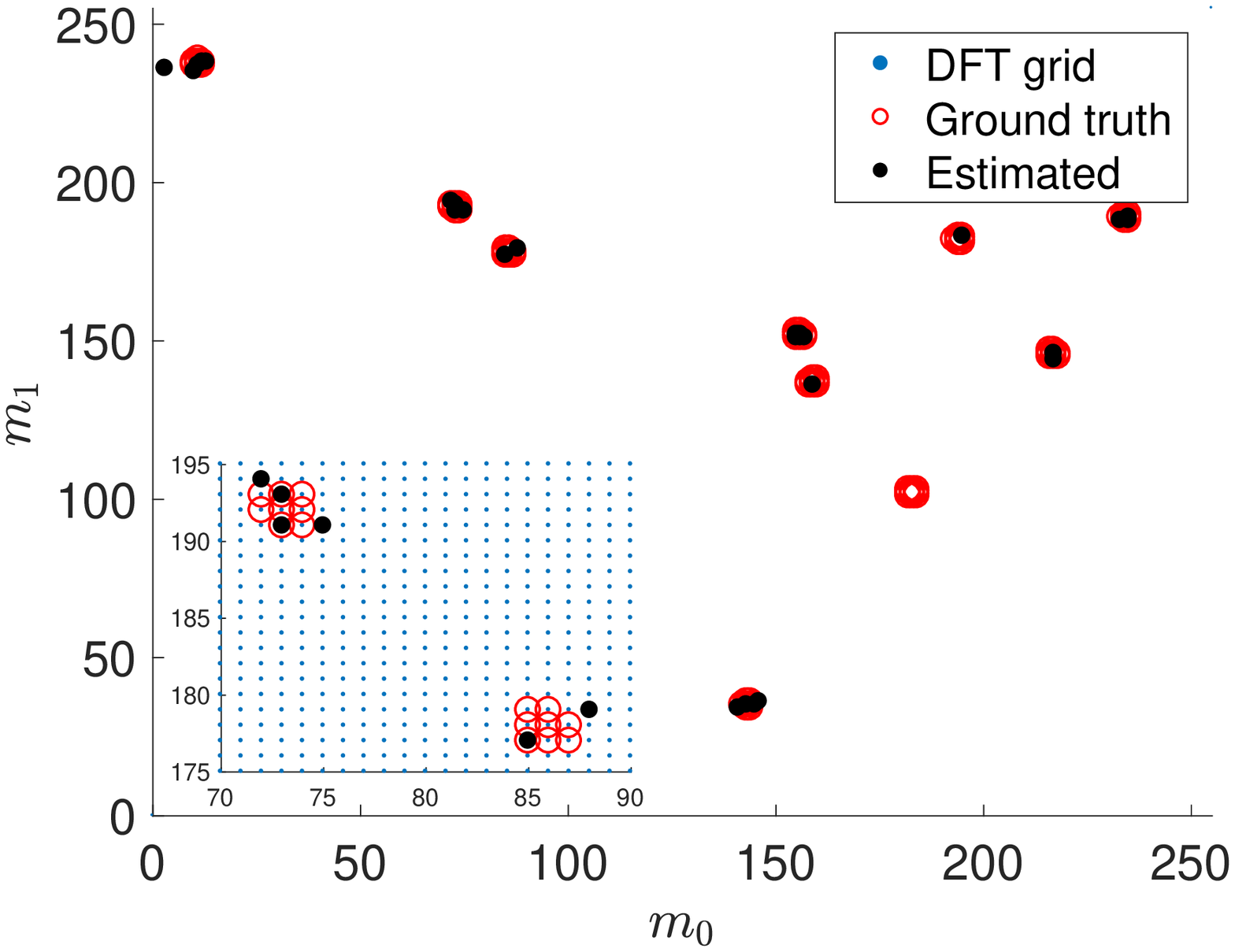} }}%
    \subfloat[]{{\includegraphics[scale=0.21]{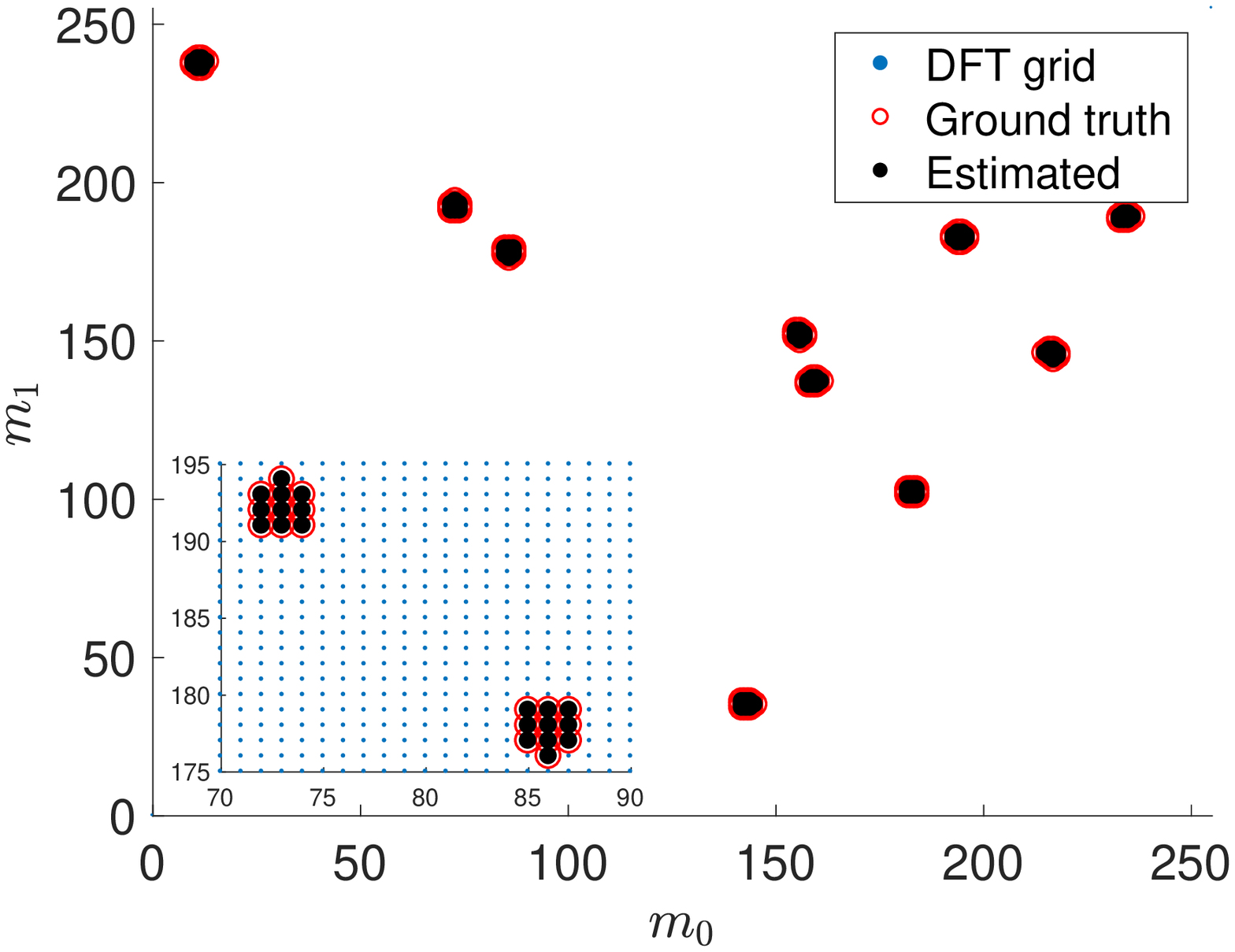} }}%
    \vskip -5pt
    \caption{$2$-D frequency recovery with different window. $K = 10,  \sigma_n = 1, a_{min} = a_{max}, SNR_{max} = 30dB, (n_s,n_d) = (3,2), T = 30$. Dolph-Chebyshev windows with various PSR are adopted. The ground truth represents (\ref{eq:sigModel2}), which relates to the window. A windows with insufficient PSR leads to miss detections and false alarms, while a  window with sufficiently large PSR yields good performance in frequency localization, albeit resulting into larger frequency cluster size.  (a) $\rho_w = 45dB$. (b) $\rho_w = 70dB$.}
    \label{fig:winComp}
\end{figure}

Fig. \ref{fig:winComp} demonstrates localization of off-grid $2$-D frequencies of RFPS-SFT using Dolph-Chebyshev window for various values of PSR. A windows with insufficient PSR leads to miss detections and false alarms (see Fig. \ref{fig:winComp} (a)), while a window with sufficiently large PSR yields good performance in frequency localization, with a trade-off of causing larger frequency cluster sizes (see Fig. \ref{fig:winComp} (b)). The ground truth in Fig. \ref{fig:winComp} represents (\ref{eq:sigModel2}), which relates to the window. 

\setlength{\belowcaptionskip}{-20pt}
\setlength{\textfloatsep}{5pt}

\setlength{\belowcaptionskip}{-20pt}
\setlength{\textfloatsep}{5pt}

\smallskip
\noindent \textbf{Effect of voting on frequency localization:}
The $n_d$-out-of-$n_s$ voting in frequency decoding procedure of RFPS-SFT can significantly reduce the false alarm rate. A low false alarm rate in each iteration of RFPS-SFT is required since the false frequencies would enter the next iteration of RFPS-SFT, which creates more false frequencies. For a fixed $n_s$, the larger the $n_d/n_s$ is, the smaller the false alarm rate is. However, this involves a trade-off between  false alarm rate and  complexity; specifically, the smaller the false alarm rate, the larger the number of the iterations required to recover all the significant frequencies. 

\begin{figure}[!htp]
    \centering
    \subfloat[]{{\includegraphics[scale=0.23]{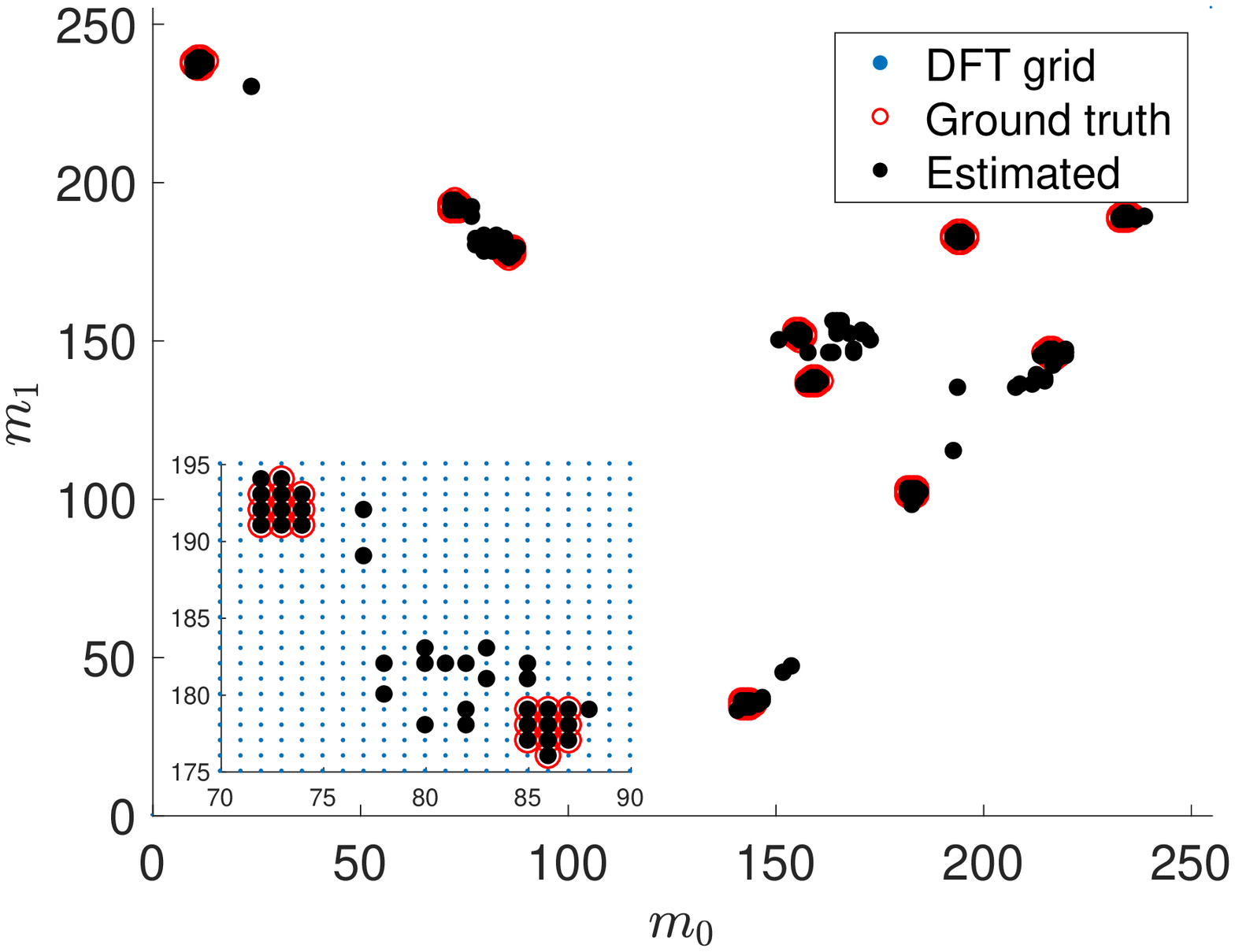} }}%
    \subfloat[]{{\includegraphics[scale=0.23]{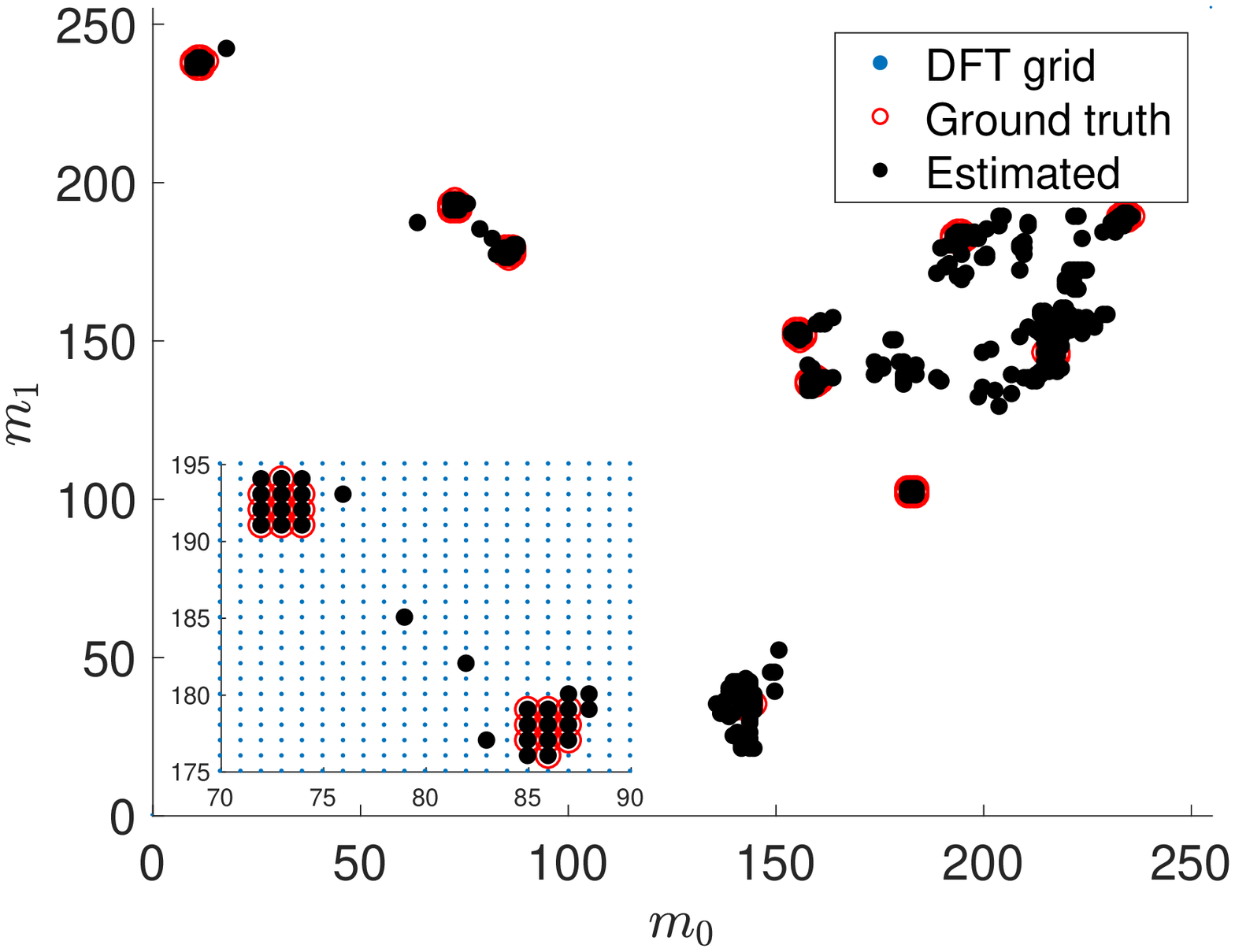} }}%
    \vskip -5pt
    \caption{Effect of voting on $2$-D frequency recovery. $K = 10,  \sigma_n = 1, a_{min} = a_{max}, SNR_{max} = 30dB$. $T = 30$ for (a)-(c). Dolph-Chebyshev windows with $\rho_w = 70 dB$ is applied. The $n_d$-out-of-$n_s$ voting procedure significantly improves frequency localization performance when $(n_d,n_s)$ is  properly designed.  (a) $(n_d,n_s) = (1,1)$. (b) $(n_d,n_s) = (3,1)$. }
    \label{fig:voting}
\end{figure}

Figs. \ref{fig:voting} and Fig. \ref{fig:winComp} (b) show the examples of $2$-D frequency recovery using different $(n_s,n_d)$. In Fig. \ref{fig:voting} (a), we set $(n_s,n_d) = (1,1)$, which reduces to the frequency localization in FPS-SFT, i.e., without voting. In this case, one can see that many false frequencies are generated. Figs. \ref{fig:voting} (b) and Fig. \ref{fig:winComp} (b) show the frequency localization result with $(n_s,n_d)$ equal to $(3,1)$ and $(3,2)$, respectively; while the former generates large amount of false frequencies, the latter exhibits ideal performance.

\begin{figure}[!htp]
    \centering
    \subfloat[]{{\includegraphics[scale=0.21]{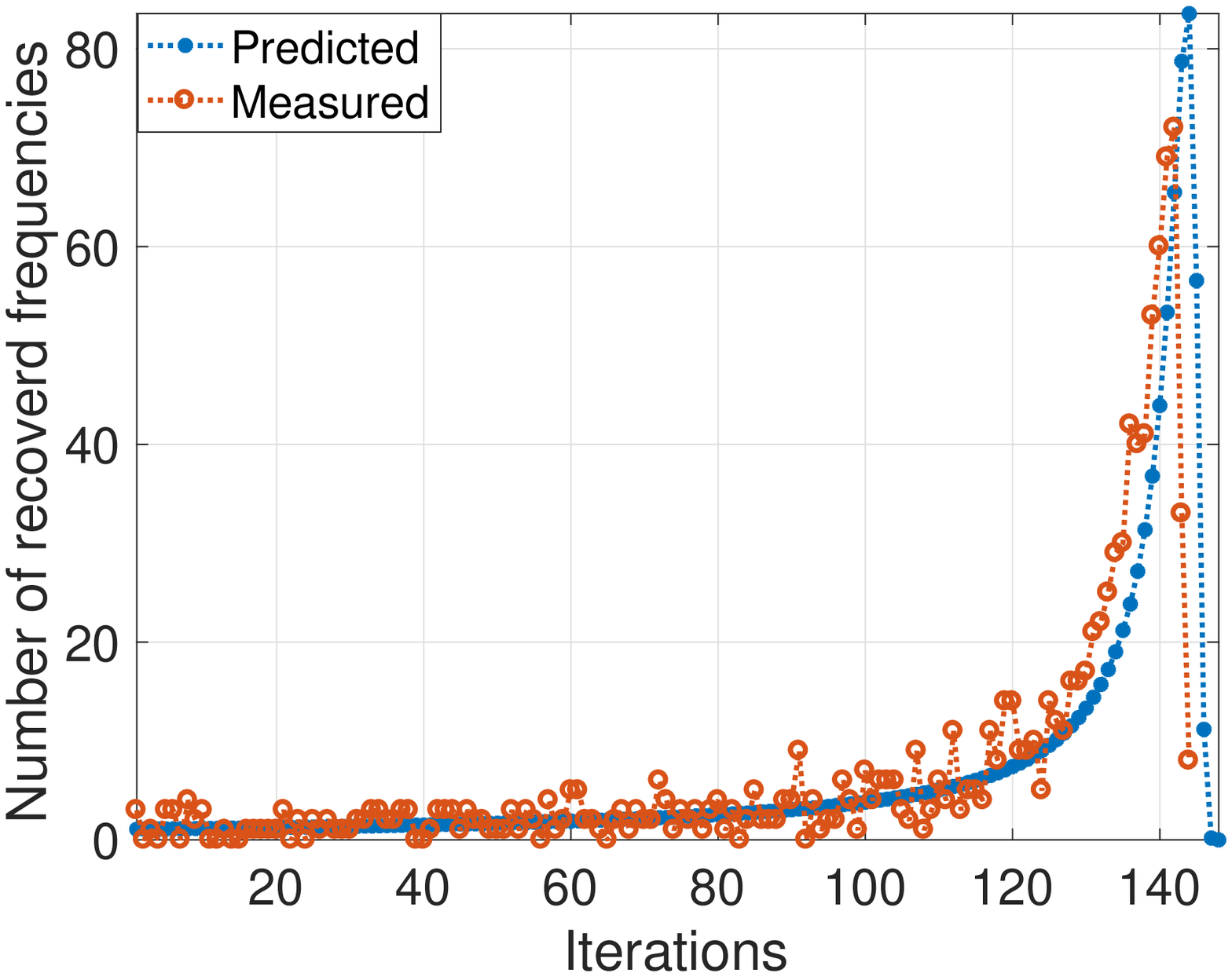} }}%
    \subfloat[]{{\includegraphics[scale=0.21]{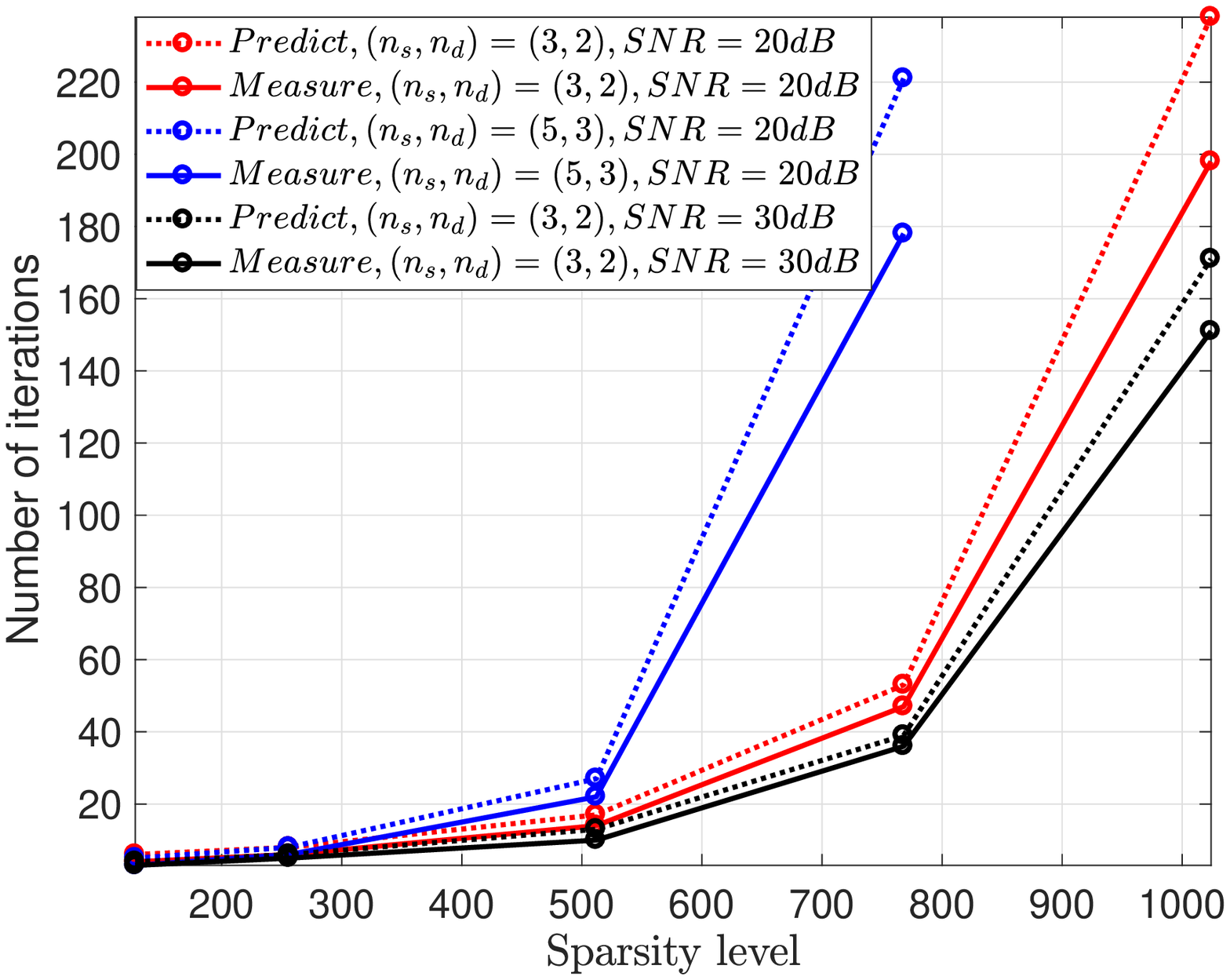} }}%
    \vskip -5pt
    \caption{Effect of SNR and sparsity level on number of iterations of RFPS-SFT.  (a) $|\b{S}|'=1000, SNR = 30dB, \sigma_p = 1/6$. (b) Comparison of predicted and measured number of iterations for various SNR and sparsity level, $|\b{S}'|$. }
    \label{fig:iterations}
\end{figure}

\smallskip
\noindent \textbf{Effect of the SNR and the sparsity level on the number of iterations of RFPS-SFT:}
The number of iterations of RFPS-SFT to recover all the significant frequencies is affected by the SNR and the sparsity level of the signal. A low SNR and less sparse signal requires large number of iterations. As discussed in Section \ref{sec:localization}, we are able to estimate the largest number of iterations that recovers $\b{S}'$. Figs. \ref{fig:iterations} (a) shows the predicted and measured number of recovered frequencies in each iteration of RFPS-SFT for $|\b{S}'|$ equal to $1000$. Fig. \ref{fig:iterations} (b) shows the predicted and  measured number of iterations of RFPS-SFT for signal with various SNR and sparsity level. The figure shows that the number of iterations upper bounds are  consistent with the measurements.      
\setlength{\belowcaptionskip}{-20pt}

\begin{figure}[!htp]
    \centering
    \subfloat[]{{\includegraphics[scale=0.22]{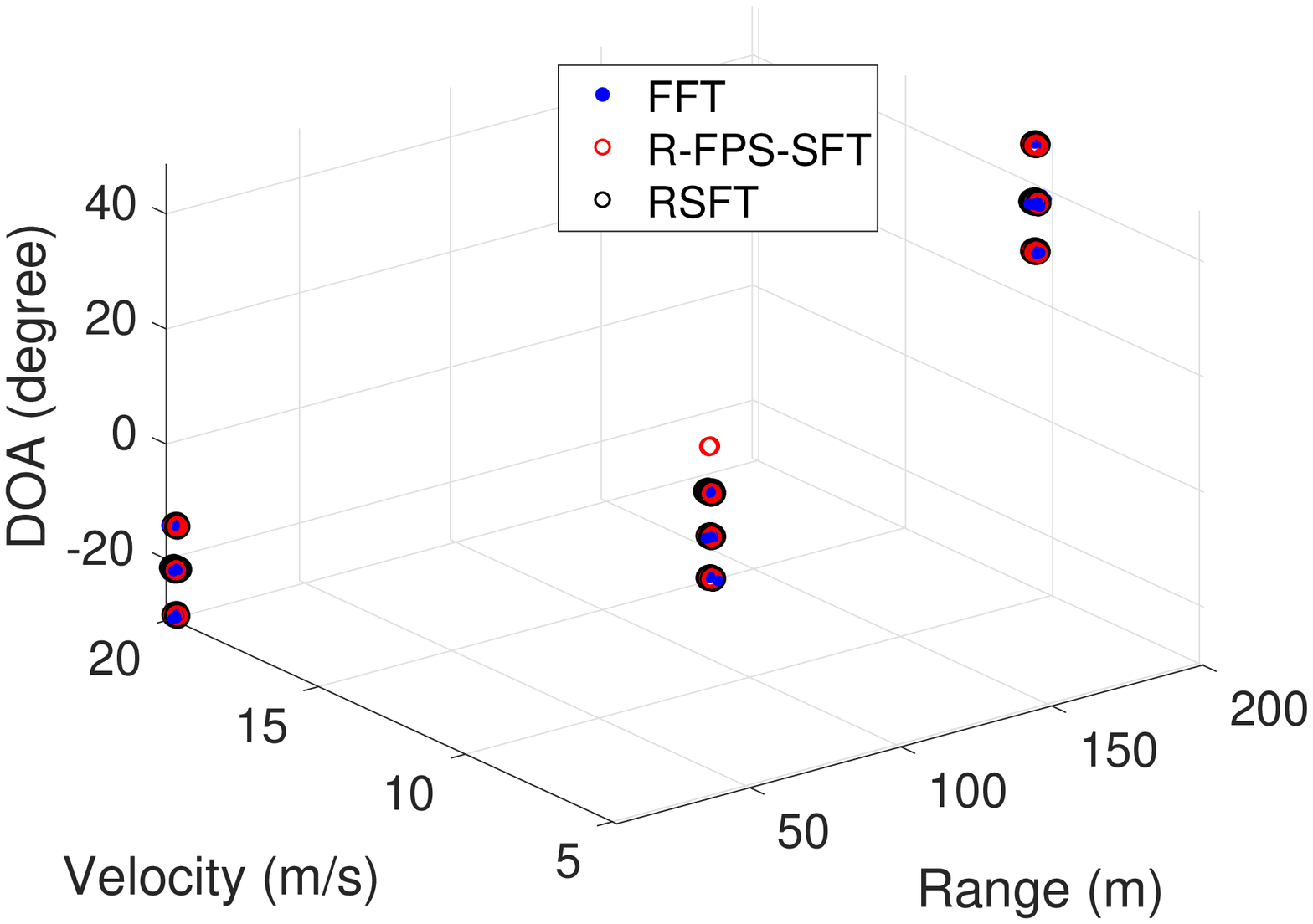} }}%
    \subfloat[]{{\includegraphics[scale=0.22]{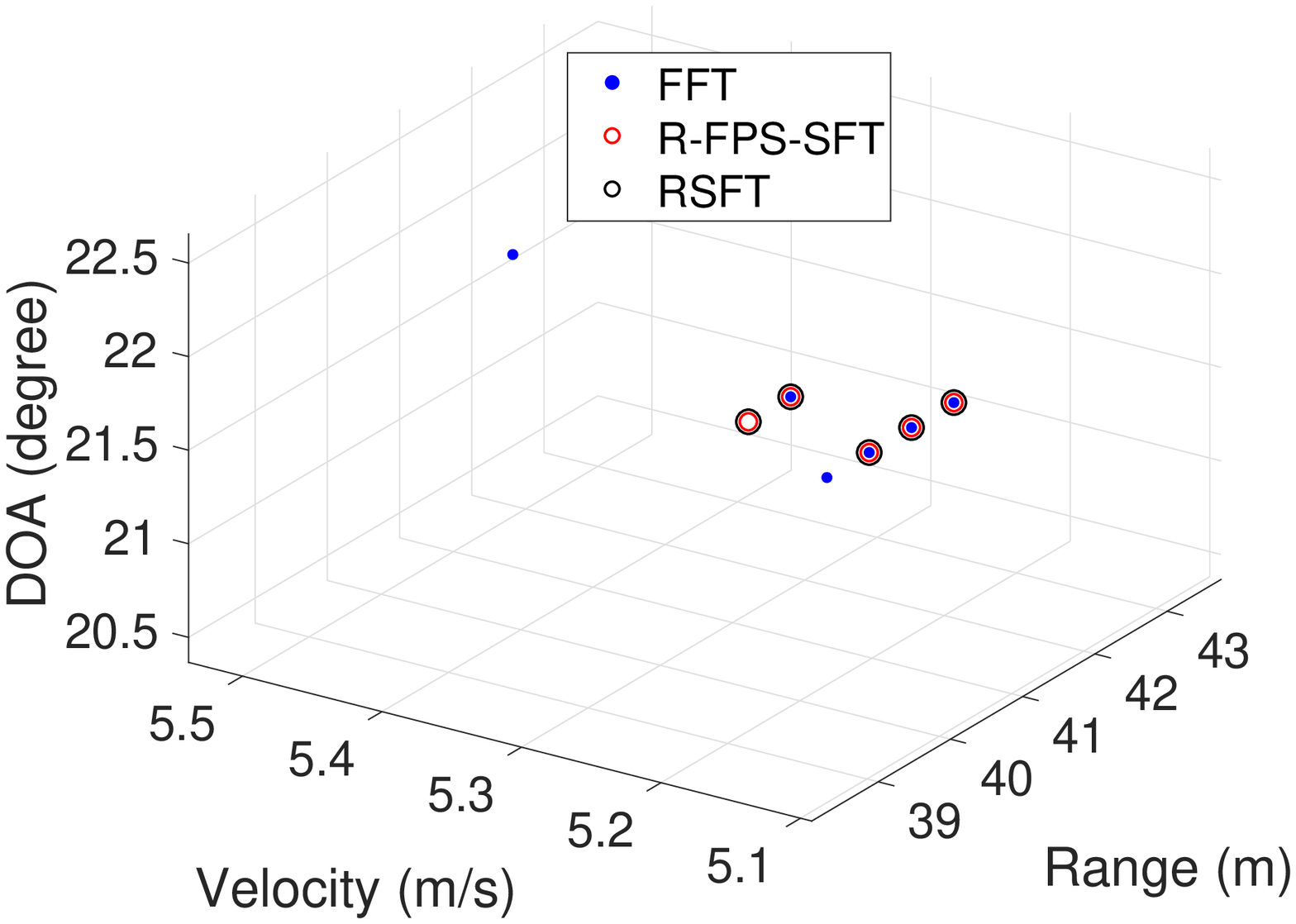} }}%
    \vskip -5pt
    \caption{Radar target reconstruction via FFT, FPS-SFT and RSFT. (a) Reconstruction of three targets. (b) Details of the frequency locations that are reconstructed for one of the three targets.}
    \label{fig:radarRec}
\end{figure}

\smallskip
\noindent \textbf{Radar target reconstruction:}
We simulate the target reconstruction for a DBF automotive radar via RFPS-SFT and compare with the RSFT. 
The main radar parameters are listed in Table \ref{tb:radar_par}; such radar configuration represents a typical long-range DBF radar \cite{engels2017advances} except that we set the number of antenna elements to be moderately large to provide a better angular measurement performance. Fig. \ref{fig:radarRec} shows the target reconstruction of $3$ radar targets via $3$-D FFT, RFPS-SFT and RSFT. All the three algorithms are able to reconstruct all the targets. Compared to the FFT and RSFT,  RFPS-SFT only requires approximately $3\%$ of data samples, which exhibits low sample complexity. However, we note that RFPS-SFT requires larger SNR than the FFT and the RSFT based methods. In near range radar applications, such as automotive radar, high SNR is relatively easy to obtain.

\begin{table}[!t] 
\caption{Radar Parameters}
\label{tb:radar_par}
\centering
\begin{tabular}{|c|c|c|}
 \hline
 \textbf{Parameter} & \textbf{Symbol} & \textbf{Value} \\
 \hline
 Center frequency & $f_c$ & $76GHz$ \\
 Pulse bandwidth & $b_w$ & $200MHz$ \\
 Pulse repetition time & $T_p$ & $89us$ \\
 Number of range bins& $N_0$& $512$  \\
 Number of PRI &$N_1$& $256$  \\
 Number of antenna elements& $N_2$ & $16$  \\
 Maximum range &$R_{max}$& $300 m$  \\
\hline
\end{tabular}
\end{table}

\section{Conclusion} \label{sec:conclusion}
In this paper, we have proposed RFPS-SFT, a robust extension of the SFT algorithm based on Fourier projection-slice theorem. We have shown that RFPS-SFT can address multidimensional data that contains off-grid frequencies and noise, while enjoys low complexity. Hence the proposed RFPS-SFT is suitable for the low-complexity implementation of multidimensional DFT based signal processing, such as the signal processing in DBF automotive radar.

\bibliographystyle{ieeetr}
\bibliography{./SFTbib}

\end{document}